\documentclass[12pt]{article}
\usepackage{amsfonts,amssymb,amsmath,graphicx,epic}

\newcommand{\N}{{\cal N}}

\newcommand{\Dslash}{D\!\!\!\!/\,}

\newcommand{\hA}{{\hat A}}

\newcommand{\hM}{{\hat M}}
\newcommand{\hN}{{\hat N}}
\newcommand{\hP}{{\hat P}}
\newcommand{\hQ}{{\hat Q}}
\newcommand{\hR}{{\hat R}}
\newcommand{\hS}{{\hat S}}

\begin{document}

\thispagestyle{empty}
\renewcommand{\thefootnote}{\fnsymbol{footnote}}

{\hfill \parbox{3.1cm}{
         HUTP-06-A0028 \\
}}

\bigskip\bigskip

\begin{center} \noindent \Large \bf
Spectroscopy of fermionic operators in AdS/CFT\\
\end{center}

\bigskip\bigskip\bigskip

\begin{large}
\centerline{  
Ingo Kirsch }
\end{large}

\bigskip\bigskip
\begin{center}

\textit{Jefferson Laboratory of Physics, Harvard University, \\
  Cambridge, MA 02138, USA}\\
\end{center}

\bigskip
\bigskip\bigskip

\bigskip\bigskip

\renewcommand{\thefootnote}{\arabic{footnote}}

\centerline{\bf \small Abstract}
\medskip

{\small We compute the spectrum of color-singlet fermionic operators
in the $\N=2$ gauge theory on intersecting D3 and D7-branes using the
AdS/CFT correspondence. The operator spectrum is found analytically by
solving the equations for the dual D7-brane fluctuations. For the
fermionic part of the D7-brane action, we use the Dirac-like form
found by Martucci {\em et al.}~(hep-th/0504041).  We also consider the
baryon spectrum of a large class of supersymmetric gauge theories
using a phenomenological approach to the gauge/gravity duality.}

\section{Introduction}

Within the AdS/CFT correspondence \cite{Maldacena:1997re}, there has
been steady progress in the dual gravity description of strongly
coupled Yang-Mills theories with matter in the fundamental
representation of the gauge group. 

One of the simplest brane configurations which realizes fundamental
matter is given by the D3/D7 brane intersection. Strings stretching
between the D3 and the D7-branes give rise to fundamental $\N=2$
hypermultiplets (``quarks'') which couple to the $\N=4$ super
Yang-Mills multiplet on the D3-branes. In the probe approximation, the
dual supergravity background corresponds to probe D7-branes embedded
into $AdS_5 \times S^5$~\cite{Karch}. Here, open string fluctuations
on the D7-branes are dual to meson operators in the $\N=2$ gauge
theory. This fluctuation-operator relation was used in
\cite{Kruczenski} to derive the meson spectrum in dependence of the
quark mass.  A similar method was applied first in \cite{Babington} to
study meson spectra and chiral symmetry breaking in confining gauge
theories. Further work on the string theory computation of meson spectra 
in strongly coupled gauge theories can be found in 
\cite{Kruczenski:2003uq}--\cite{Erdmenger:2006bg}.

While the main focus has been on the holographic study of scalar and
vector mesons, not much attention has been paid to the spectra of
operators with half-integer spin. In particular, baryons have been
widely neglected so far, although, in principle, a brane configuration
for dynamical baryons is known \cite{Kruczenski} (see also
\cite{Sakai:2004cn} for a similar proposal): A baryon vertex is
realized in string theory by $N$~fundamental strings attached to a
D5-brane which is wrapped on the five-sphere of $AdS_5 \times S^5$
\cite{Witten}.  By adding a stack of D7-branes at some finite distance
in the AdS space, the $N$ fundamental strings would end on the
D7-branes rather than on the boundary of the AdS space.  This
corresponds to $N$ dynamical quarks with a finite mass.

There are several difficulties which impede immediate progress towards
such a holographic description of dynamical baryons: i) In large-$N$
field theories baryons consist of $N$ quarks. It is not clear whether
one can neglect the backreaction of the baryonic D5-brane on the AdS 
geometry. ii) The set-up consists of several different branes and
the topology of the baryon vertex is quite complex. iii) Baryons have
half-integer spin and the dual fluctuations will be described by
Dirac equations, Rarita-Schwinger equations, etc., all on curved
spacetimes.

In this paper we will not directly approach baryons. Instead, we will
discuss the spectrum of certain fermionic operators in the D3/D7
system. As the baryons, these operators have half-integer spin, but
do not require the introduction of an additional brane in the dual
string theory set-up. We may therefore develop techniques for dealing
with fermionic open string fluctuations, while avoiding technical
difficulties connected with a baryonic D5-brane.

The operators we are interested in are the supersymmetric partners of
the meson-like operators discussed by Kruczenski~{\it et
al.}~\cite{Kruczenski}.  There are two towers of spin-$\frac{1}{2}$
operators: one with the dimension-$\frac{5}2$ operator ${\cal F} \sim
\tilde \psi q$, the other with the dimension-$\frac{9}2$ operator
${\cal G} \sim \tilde \psi \lambda \psi$ at the bottom of the
tower. Here $q$, $\tilde q$ and $\psi$, $\tilde \psi$ denote
fundamental scalars and spinors. The operator $\tilde \psi \lambda
\psi$ contains an additional adjoint spinor $\lambda$.

Similarly to the case of the scalar and vector mesons, we obtain the
fermionic operator spectrum by considering D7-brane fluctuations dual
to the above operators. Since the spin of the fluctuations must be
identical to that of the operators, we start from the {\em fermionic}
part of the D7-brane action as constructed in \cite{Martucci}. The
corresponding equation of motion effectively describes a Dirac spinor on
$AdS_5 \times S^3$ which couples to the self-dual five-form flux of
type IIB string theory. The Kaluza-Klein reduction on the three-sphere
$S^3$ then provides two sets of modes $\Psi^\pm_\ell$ with masses
$m^\pm_\ell$ ($\ell=0,1,2,...$) satisfying the mass-dimension relation
for spin-$\frac{1}{2}$ fields \cite{Henningson}:
\begin{align}
\Delta_\ell=|m^\pm_\ell|+2 \,, \label{eqn1}
\end{align}
where $\Delta_\ell$ denotes the conformal dimensions of the operators
${\cal F}^\ell$ and ${\cal G}^\ell$, the higher-$\ell$ relatives of
${\cal F}$ and ${\cal G}$.  Since also the quantum numbers of these
operators match exactly those of $\Psi^\pm_\ell$, this ensures the
correct fluctuation-operator map between $\Psi^\pm_\ell$ and ${\cal
F}^\ell$, ${\cal G}^\ell$.  After recasting the equations of motion
into a second order form, we find the spectrum of ${\cal F}^\ell$ and
${\cal G}^\ell$ by assuming a plane-wave behavior of the fluctuations
$\Psi^\pm_\ell$. The resulting spectrum is linear in the quark mass and
agrees exactly with the spectrum expected from supersymmetry.

We finally come back to baryons in the last section of the paper in
which we construct a phenomenological supergravity model for baryonic
operators in a large class of supersymmetric gauge theories. Instead
of starting from a ten-dimensional brane configuration, we fix the
background from the properties of the dual field theory. This is known
as the ``bottom-up'' approach to AdS/CFT \cite{Erlich, DaRold:2005vr,
Teramond2004} and has first been applied to baryons by Brodsky and
Teramond \cite{Teramond2004}.  As in \cite{Teramond2004}, we assume
that fluctuations dual to spin-$\frac{1}{2}$ baryons are effectively
described by a massive Dirac equation on $AdS_5$. The mass of the
Dirac spinor is fixed by the conformal dimension of the baryon
operator. Deviating from \cite{Teramond2004}, we consider gauge
theories which are superconformal in the UV, at least in some
parameter regime. The baryon spectrum will therefore be a function of
the quark mass rather than the infra-red cut-off $\Lambda_{QCD}$. We
fill find that the large-$N$ baryon spectrum is linear in the quark
mass and scales with $N$ as expected from field theory
\cite{Witten:1979kh}.

The paper is organized as follows. In Sec.~\ref{sec2} we review both
the low-energy effective field theory of the D3/D7 intersection and
the holographic computation of meson spectra. In Sec.~\ref{sec3} we
adapt the holographic method to find the spectrum of operators with
half-integer spin. In Sec.~\ref{sec4} we discuss large $N$ baryons in
an effective approach to the gauge/gravity duality.  In Sec.~5 we
briefly summarize our results and discuss open problems.

\section{Spectroscopy of meson operators in AdS/CFT}
\label{sec2}

In the following we briefly discuss the $\N=2$ world-volume theory of
the D3/D7 intersection and review the holographic computation of its
meson spectrum \cite{Kruczenski}. Those who are familiar with the meson
spectroscopy in the D3/D7 theory may wish to proceed immediately to
Sec.~\ref{sec3} in which we adapt the method to compute the spectrum
of fermionic operators.

\subsection{The D3/D7 brane intersection} \label{sec21}

The D3/D7 brane intersection in flat space consists of a stack of
$N_c$ coincident D3-branes (along 0123) which is embedded into the
world volume of $N_f$ D7-branes (along 01234567). This system
preserves $1/4$ of the total amount of supersymmetry in type IIB
string theory and has an $SO(4) \times SO(2)$ isometry in the
directions transverse to the D3-branes. The $SO(4)$ rotates in $x^4,
x^5, x^6, x^7$, while the $SO(2)$ group acts on $x^8,x^9$. Note that
separating the D3-branes from the D7-branes in the 89 direction by a
distance $L$ explicitly breaks the $SO(2)$ group.

The world-volume theory describes a four-dimensional $\N=4$ super
Yang-Mills multiplet which is coupled to $N_f$ $\N=2$ hypermultiplets
in the fundamental representation of the $U(N_c)$ gauge group.  Under
$\N=1$ supersymmetry the $\N=4$ vector multiplet decomposes
into the vector multiplet $W_\alpha$ and the three chiral superfields
$\Phi_1$, $\Phi_2$, $\Phi_3$. The $\N=2$ fundamental hypermultiplets
can be written in terms of the $\N=1$ chiral multiplets $Q^r,
\tilde Q_r$ ($r=1,...,N_f$).  In $\N=1$ superspace notation, the
Lagrangian is thus given by
\begin{align}
{\cal L} = {\rm Im} &\left[ \tau \int d^4 \theta  
\left( {\rm tr\,} ( \bar \Phi_I e^V \Phi_Ie^ {-V} )+ 
Q^\dagger_{r} e^V Q^r + \tilde Q^\dagger_{r}  e^{-V} \tilde 
Q^r \right) \right. \nonumber\\
&\,\,+ \left.\tau \int d^2 \theta ( {\rm tr\,}(W^\alpha W_\alpha) + W ) + c.c.
\right]\,,
\end{align}
where the superpotential $W$ is
\begin{align} 
W={\rm tr\,} (\varepsilon_{IJK} \Phi_I\Phi_J\Phi_K)
+\tilde Q_{r} (m+\Phi_3) Q^r  \,.
\end{align}

The $SO(2) \simeq U(1)$ isometry corresponds to a $U(1)_R$ R-symmetry
in the field theory which is explicitly broken by a quark mass $m_q
\sim L$. The field theory has also a global $SO(4) \approx SU(2)_\Phi
\times SU(2)_{\cal R}$ symmetry which consists of a $SU(2)_\Phi$
symmetry and a $\N=2$ $SU(2)_{\cal R}$ R-symmetry. The global symmetry
$SU(2)_\Phi$ rotates the scalars in the adjoint hypermultiplet. There
is also a group $U(1)_B$ which is a subgroup of the $U(N_f)$ flavor
group. The fundamental superfields $Q^r$ ($\tilde Q_r$) are charged
$+1$ ($-1$) under $U(1)_B$, while the adjoint fields are inert.  The
components of the $\N=1$ superfields and their quantum numbers are
summarized in the Tab.~\ref{tablefields} (see also \cite{Strassler})
which we will need for the construction of operators.

\begin{table}
\begin{center}
\begin{tabular} {|c|c|c|c|c|c|c|c|}
\hline
&components & spin & $SU(2)_\Phi \times SU(2)_{\cal R}$ & $U(1)_{\cal R}$ &$\Delta$ & $U(N_f)$ & $U(1)_B$\\
\hline
$\Phi_1, \Phi_2$ & $X^4, X^5, X^6,  X^7$ &$0$& $(\frac{1}{2},\frac{1}{2})$& $0$ & $1$& 1 & 0\\
 &$\lambda_1, \lambda_2$ & $\frac 1 2$ & $(\frac{1}{2},0)$& $-1$ & $\frac{3}{2}$& 1 & 0\\
\hline
$\Phi_3$, $W_\alpha$ & $X_V^A=(X^8, X^9)$ & $0$ &$(0,0)$ & $+ 2$ &$1$& 1 & 0 \\
&$\lambda_3, \lambda_4$ &$\frac 1 2$& $(0,\frac{1}{2})$& $+1$&$\frac{3}{2}$ & 1
& 0\\
&$v_\mu$& $1$ &$(0,0)$ & $0$& $1$& 1& 0\\
\hline
$Q$, $\tilde Q$ & $q^m=(q, \bar {\tilde q})$&$0$ & $(0,\frac{1}{2})$& $0$&$1$& $N_f$ & +1\\
&$\psi_i=(\psi, \tilde \psi^\dagger)$ & $\frac{1}{2}$ & $(0,0)$& $\mp 1$& $\frac{3}{2}$ &$N_f$&+1 \\
\hline
\end{tabular}
\end{center}
\caption{Fields of the D3/D7 low-energy effective field theory and
their quantum numbers under the global symmetries. Note that
$U(1)_B \subset U(N_f)$.}
\label{tablefields}
\end{table}

The exact perturbative $\N=2$ beta function for the 't~Hooft coupling
is proportional to $N_f/N_c$ \cite{Strassler, Vaman}. In the probe
approximation, the D7-branes do not backreact on the $AdS_5 \times
S^5$ near-horizon geometry of the D3-branes and the field theory is
conformal corresponding to the strict $N_f/N_c \rightarrow 0$
limit. Beyond the probe approximation, $N_f/N_c$ is finite, and the
beta function is positive, implying an UV Landau pole in the field
theory. This is a pathology of the perturbative field theory and is
reflected by a dilaton divergence in the fully localized supergravity
solution of the D3/D7 setup \cite{Aharony, Grana,
Burrington, DiVecchia2, Vaman}.  Despite the occurrence of logarithmic
tadpoles, this background is still consistent as far as the embedding
of a nonconformal field theory is concerned, see Ref.~\cite{Vaman,
DiVecchia2, DiVecchia} and references therein. Note that logarithmic
tadpoles do not represent gauge anomalies, but instead provide the
correct one-loop running of the gauge coupling. The supergravity
background perfectly reflects the properties of the perturbative field
theory.

\subsection{Review on meson spectroscopy in the D3/D7 set-up}
\label{sec22}

The $\N=2$ world-volume field theory of the D3/D7 brane intersection
contains the scalar meson operators
\begin{align} \label{mesonop}
 {\cal M}^{A\ell}_{s} = \bar \psi_i \sigma^A_{ij} X^\ell
 \psi_j + \bar q^m X^A_V X^\ell q^m \,\qquad(i, m=1,2)
\end{align}
which have conformal dimensions $\Delta=3+\ell$. Here $X^A_V$ denotes
the vector $(X^8, X^9)$ and $\sigma^A=(\sigma^1, \sigma^2)$ is a
doublet of Pauli matrices. Both $X^A_V$ and $\sigma^A$ transform in
the ${\mathbf 2}$ of $U(1)_R$. The components $q^m$ and $\psi_i$ of
the fundamental hypermultiplets are as in Tab.~\ref{tablefields}.
$X^\ell$ denotes the symmetric, traceless operator insertion $X^{\{
i_1} \cdots X^{ i_\ell \}}$ of $\ell$ adjoint scalars $X^i$
($i=4,5,6,7$) which transform in the fundamental representation
$(\frac{1}{2}, \frac{1}{2})$ of $SO(4) \approx SU(2)_\Phi \times
SU(2)_{\cal R}$. The operators ${\cal M}^{A\ell}_{s}$ thus transform
in the $(\frac{\ell}{2}, \frac{\ell}{2})$ of $SO(4)$ and are charged
$+2$ under $U(1)_R$.

The spectrum of these operators (and that of the vector mesons) was
found in \cite{Kruczenski} by evaluating the Dirac-Born-Infeld (DBI)
action for a probe D7-brane. The D7-brane wraps an $AdS_5 \times S^3$
submanifold inside the $AdS_5 \times S^5$ near-horizon geometry of the 
D3-branes. In static gauge, where the
world volume coordinates of the D7 brane are identified with the
spacetime coordinates by $\xi^a \sim t,x_1,..., x_7$,
the DBI action is given by~\cite{Kruczenski}
\begin{align} \label{DBI}
S^{b}_{D7} & =  - T_7 \int d^8 \xi \sqrt{-\det g_{ab}^{PB}} \nonumber\\
&=- T_7 \int d^8 \xi ~ \epsilon_3 ~ \rho^3
\sqrt{1 + {\frac{g^{ab}}{  \rho^2 + x_8^2 + x_9^2}}(\partial_a x_8
\partial_b x_8 +
\partial_a x_9 \partial_b x_9)} \,,
\end{align}
where $g_{ab}^{PB}$ is  the pullback of the $AdS_5  \times S^5$ metric
on the D7 world-volume.  Moreover, $g_{ab}$ denotes the induced metric
on the  D7 brane and $\epsilon_3$  is the determinant  factor from the
three sphere.  $x_8$  and $x_9$ are the coordinates  transverse to the
D7-brane, while $\rho^2=x_4^2+...+ x^2_7$.

One easily verifies \cite{Kruczenski} that e.g.\ $x_8=0$, $x_9=L$ is a
solution to the corresponding Euler-Lagrange equations. The constant
$L$ is the distance between the D3- and the D7-branes and is
proportional to the mass of the fundamental hypermultiplets
(``quarks''). The spectrum of the operators ${\cal M}^\ell_{s}$ can
then be found by considering fluctuations of the plane-wave type
around this ground state solution. More precisely, one makes the
ansatz
\begin{align}
x_8 = 0 \,, \qquad x_9=L+f_\ell(\rho) e^{ik \cdot x} {\cal Y}^\ell(S^3) \,,
\end{align}
where $M^2=-k^2$ is interpreted as the meson mass. The functions
${\cal Y}^\ell(S^3)$ are the scalar spherical harmonics on $S^3$ which
have eigenvalues $-\ell(\ell+2)$ and transform in the $(\frac{\ell}2,
\frac{\ell}2)$ of $SO(4)$. Substituting the ansatz into the
Euler-Lagrange equations obtained from the DBI action (\ref{DBI})
expanded to quadratic order in the fields, one obtains the following
equation for the fluctuations $f_\ell(\rho)$:
\begin{align} \label{mesoneqnKru}
\partial_\rho^2 f_\ell(\rho)+ \frac{3}{\rho} \partial_\rho  f_\ell(\rho)+
\left(\frac{M^2}{(\rho^2+L^2)^2} - \frac{\ell(\ell+2)}{\rho^2}\right)
f_\ell(\rho) =0\,.
\end{align}
This equation is solved in terms of the hypergeometric function
by \cite{Kruczenski}
\begin{align}
f_\ell(\rho) = \frac{\rho^\ell}{(\rho^2+L^2)^{n+\ell+1}} 
F(-(n+\ell+1), -n; \ell+2; -\rho^2/L^2)\,,
\end{align}
where the excitation number $n$ is related to the scalar meson 
mass by
\begin{align}
 M^2_{s}=\frac{4L^2}{R^4}(n+\ell+1)(n+\ell+2)  \qquad (n, \ell \geqslant 0)\,.
\end{align}
The discreteness of the spectrum follows from the normalizability of
the states. Note that the spectrum is a linear function of the quark
mass $m_q \sim L$. This is a particular feature of the superconformal
field theory. It has been shown in \cite{Babington} (see also
\cite{Kruczenski:2003uq}) that in a nonsupersymmetric deformation of
the D3/D7 world-volume theory, the meson mass satisfies the
Gell-Mann-Oakes-Renner-relation for small quark masses.

\section{Spectroscopy of fermionic operators}
\label{sec3}

\subsection{Spectroscopy of spin-$\frac{1}{2}$ operators}
\label{sec31}

We now use the holographic method to derive the spectrum of certain
fermionic operators in the D3/D7 theory. More specifically, the
operators we are interested in are the supersymmetric partners of the
meson-like operators studied in \cite{Kruczenski}.  There are two
classes of spin-$\frac{1}{2}$ operators:
\begin{align} \label{opF}
{\cal F}^{\ell}_{\alpha} &\sim \bar q 
 X^{\ell} \tilde \psi^\dagger_\alpha  + \tilde \psi_\alpha X^{\ell} q  \,, \\
 {\cal G}^\ell_{\alpha } &
 \sim \bar \psi_i \sigma^B_{ij} 
 \lambda_{\alpha C} X^\ell \psi_j 
 + \bar q^m X_V^B \lambda_{\alpha C} X^\ell q^m
 \,,  \qquad(A,B,C=1,2) 
\end{align}
which have the conformal dimensions $\Delta=\frac{5}{2}+\ell$ and
$\Delta=\frac{9}{2}+\ell$ $(\ell \geqslant 0)$, respectively. Both
types of operators have fundamental fields at their ends: scalars
$q^m=(q, \bar {\tilde q})^T$ and spinors $\psi_i=(\psi, \tilde
\psi^\dagger)^T$.  As in Eq.~(\ref{mesonop}), the operator insertion
$X^\ell$ generates operators with higher angular momentum $\ell$.  The
spinors $\lambda_{\alpha A}$ ($A=1,2$) belong to the adjoint
hypermultiplets $(\Phi_1, \Phi_2)$. The quantum numbers of these
fields are listed in Tab.~\ref{tablefields}.

The operators ${\cal F}^\ell_\alpha$ and ${\cal G}^\ell_\alpha$ have
the following quantum numbers under the global symmetries of the
theory: Since $q^m$, $\psi^i$ and $X^\ell$ have the $SO(4)$ quantum
numbers $(0, \frac{1}{2})$, $(0,0)$ and $(\frac \ell 2, \frac \ell
2)$, respectively, ${\cal F}^\ell_\alpha$ transforms in the
$(\frac{\ell}{2},\frac{\ell+1}{2})$ of $SO(4) \approx SU(2)_\Phi
\times SU(2)_{\cal R}$. The $U(1)_R$ charge of ${\cal F}^\ell_\alpha$
is $+1$. Similar operators can be found in the world-volume theory of
intersecting D3/D5 branes \cite{0111135}.  The operators ${\cal
G}^\ell_{\alpha}$ are obtained by inserting a doublet of adjoint
spinors $\lambda_{\alpha A}$ ($A=1,2$) into the operators ${\cal
M}^\ell_{s}$. Since $\lambda_{\alpha A}$ has the $SO(4)$ quantum
numbers $(\frac{1}{2},0)$, the operators ${\cal G}^\ell_\alpha$
transform in the $(\frac{\ell+1}{2},\frac{\ell}{2})$ representations
of $SO(4)$. The $U(1)_R$ charge of ${\cal G}^{\ell}_{\alpha}$ ($+1$)
is the same as for the operators ${\cal F}^{\ell}_{\alpha}$.

Let us consider the properties of the bulk modes $\Psi^+_\ell$ and
$\Psi^-_\ell$ dual to ${\cal G}^\ell_\alpha$ and ${\cal
F}^\ell_\alpha$. The spin of $\Psi^\pm_\ell$ must be identical to that
of the boundary states, {\em i.e.}\ their spin must be $\frac{1}{2}$.
For spin-$\frac 1 2$ modes, the relation between the conformal
dimension of the field theory operator and the mass of the dual mode
in $AdS_{5}$ is given by Eq.~(\ref{eqn1}). The $AdS_5$ modes
$\Psi_\ell^\pm$ should therefore have the masses
\begin{align} \label{masses}
|m_\ell^+|= \frac{5}{2}+\ell \,,\qquad 
|m_\ell^-|= \frac{1}{2}+\ell \,.
\end{align}
Moreover, ${\cal F}^\ell_\alpha$ and ${\cal G}^\ell_\alpha$ have
fundamental fields at their ends, for which reason the dual bulk modes
$\Psi^\pm_\ell$ must descend from open string fluctuations.  Recall
that, since the D7-brane does not backreact on the geometry, closed
strings are only dual to pure adjoint operators. In other words, the
modes must again correspond to open string fluctuations on the
D7-brane. However, since the operators have half-integer spin, the D7
fluctuations must be fermionic.  We now show that the D3/D7 brane
configuration indeed contains spin-$\frac 1 2$ open string modes with
these properties.

For this, we consider the fermionic part of the D7-brane action which
to quadratic order in the fermions is given by \cite{Martucci}
\begin{align} \label{fermionicDBI}
S^{f}_{D7}= \frac{\tau_{D7}}{2} \int d^{8}\xi \sqrt{-\det g} \hat{\bar
   \Psi} {\cal P}_- \Gamma^\hA ( D_\hA + \frac{1}{8} 
   \frac{i}{2 \cdot 5!} F_{\hN\hP\hQ\hR\hS}
   \Gamma^{\hN\hP\hQ\hR\hS} \Gamma_\hA ) \hat \Psi  \,.
\end{align}
Here $\xi^\hA$ are the world-volume coordinates ($\hA =0,...,7$)
which, in static gauge, will be identified with the spacetime
coordinates $t, x^1,...,x^7$.  The field $\hat \Psi$ is the 10d
positive chirality Majorana-Weyl spinor of type IIB string theory and
$\Gamma_\hA$ is the pullback of the 10d gamma matrix $\Gamma_{\hat M}$
($\hM, \hN, ...  =0,...,9$), $\Gamma_\hA=\Gamma_{\hat M} \partial_\hA
x^{\hat M}$.  The integration goes over the world-volume of the
D7-brane which wraps an $AdS_5 \times S^3$ submanifold of \mbox{$AdS_5
\times S^5$}.  The spinor $\hat \Psi =\hat \Psi (x^M, \theta^m)$
depends on the coordinates $x^M$ of $AdS_5$ and the three angles
$\theta^m=(\theta^1, \theta^2, \theta^3)$ of the three-sphere
$S^3$. The D7-brane is located at $\theta^4= \theta^5=0$ corresponding
to massless quarks in the field theory.  The operator ${\cal P}_-$ is
a $\kappa$-symmetry projector ensuring $\kappa$-symmetry invariance of
the action. The action is therefore invariant under supersymmetries
corresponding to any bulk Killing spinor.

We now decompose every ten-dimensional field or gamma matrix into
parts associated with $AdS_5$ and $S^5$, respectively.  Choosing a
local Lorentz frame, the 10d gamma matrices $\Gamma^{\hat M}$
decompose as
\begin{align} 
\Gamma^M &= \sigma_y \otimes \mathbf{1}_4 \otimes \gamma^M \,,\quad 
(M=0,1,2,3,4) \,,\nonumber\\
\Gamma^m &= \sigma_x \otimes \gamma^m \otimes \mathbf{1}_4\,, \quad 
(m=5,6,7,8,9) \,, \label{decomp1}
\end{align}
where $\mathbf{1}_4$ is the 4d unit matrix and $\sigma_x, \sigma_y,
\sigma_z$ are the Pauli matrices. The five-dimensional Minkowski and
Euclidean gamma matrices, $\gamma^A$ and
$\gamma^a$, satisfy the relations
\begin{align}
\{ \gamma^M, \gamma^N \} = 2 \eta^{MN} \,,\qquad
\{ \gamma^m, \gamma^n \} = 2 \delta^{mn} \,.
\end{align}
From this, one obtains
\begin{align}
\Gamma^{11} = \sigma_z \otimes \mathbf{1}_4\otimes \mathbf{1}_4 \,,\qquad
\Gamma^{01234} =  i\sigma_y \otimes \mathbf{1}_4 \otimes \mathbf{1}_4\,, 
\qquad
\Gamma^{56789} =  \sigma_x \otimes \mathbf{1}_4 \otimes \mathbf{1}_4\,.
\end{align}

The 10d spinor $\hat \Psi$ has positive chirality, $\Gamma^{11}\hat \Psi
=\hat \Psi$, and can be decomposed as
\begin{align} \label{decomp2}
\hat \Psi = \,\uparrow  \otimes \,\chi \otimes \Psi \,,
\end{align} 
where $\uparrow$ denotes the two-component spinor $(1,0)^T$, and
$\chi$ and $\Psi$ are four-component spinors of $SO(5)$ and $SO(1,4)$,
respectively. These groups act on the tangent spaces of $S^5$
and $AdS_5$, respectively. The spinor $\chi=\chi_{||} \otimes
\chi_{\perp}$ splits into a 3d spinor $\chi_{||}$ associated
with $S^3$ and a 2d spinor $\chi_{\perp}$ acting transverse to the
$S^3$.

The action (\ref{fermionicDBI}) leads to the Dirac equation
\begin{align} \label{Diraceqn}
\Dslash  \hat \Psi + \frac{1}{8}\frac{i}{2\cdot 5!} 
\Gamma^\hA F_{\hN\hP\hQ\hR\hS}
\Gamma^{\hN\hP\hQ\hR\hS} \Gamma_\hA   \hat \Psi = 0 \,,
\end{align}
where $\Dslash$ is the Dirac-operator on $AdS_5 \times S^3$. The
second term describes the coupling of $\hat \Psi$ to the self-dual
five-form field strength $F_{\hN\hP\hQ\hR\hS}$. We parametrize the 
five-form as
\begin{align} \label{5form}
F_{NPQRS} &= \frac{4}{R}\, \varepsilon_{NPQRS} \,, \quad (N, P,...=0,1,2,3,4)
\,,\nonumber\\
F_{npqrs} &= \frac{4}{R}\, \varepsilon_{npqrs} \,, \quad (n, p,...=5,6,7,8,9)
\,,
\end{align}
where $R$ is the AdS radius.

Using the decompositions (\ref{decomp1}), (\ref{decomp2})
and (\ref{5form}),
we find
\begin{align}
\frac{1}{8} \frac{i}{2\cdot 5!} \Gamma^\hA F_{\hN\hP\hQ\hR\hS} 
\Gamma^{\hN\hP\hQ\hR\hS} \Gamma_\hA \hat \Psi
= \frac{i}{16R} \Gamma^\hA \left((\sigma_x + i \sigma_y) \otimes 
\mathbf{1}_4 \otimes \mathbf{1}_4 \right) \Gamma_\hA  \hat \Psi 
= \frac{-i}{R} (\downarrow \otimes \chi \otimes \Psi)
\end{align}
and ($A=0,...,4$, $a=5,6,7$)
\begin{align}
\Dslash \hat \Psi&=\Gamma^\hA D_\hA \hat \Psi = \Gamma^A D_A \hat \Psi
+ \Gamma^a D_a \hat \Psi \nonumber\\
&=(\sigma_y  \otimes \mathbf{1}_4 \otimes \gamma^A D_A +
\sigma_x  \otimes \gamma^a D_a \otimes \mathbf{1}_4 )(\uparrow \otimes 
\chi \otimes \Psi) \nonumber\\
&= \left(i (\mathbf{1}_2 \otimes \mathbf{1}_4 \otimes
\gamma^A D_A) + (\mathbf{1}_2 \otimes \gamma^a D_a \otimes \mathbf{1}_4)
\right) (\downarrow \otimes \chi \otimes \Psi) \nonumber\\
&\equiv (i \Dslash_{AdS_5} + \Dslash_{S^3}) (\downarrow \otimes \chi \otimes 
\Psi) \,,
\end{align} 
where $\Dslash_{AdS_5}$ and $\Dslash_{S^3}$ are the Dirac operators on
$AdS_5$ and $S^3$, respectively. 

The Dirac operator $\Dslash_{S^n}$ on a $n$-sphere $S^n$ of radius $R$
and its eigenvalues $\lambda$ are well-known, see
e.g.~\cite{Camporesi}. The spinor spherical harmonics $\chi_\ell^\pm$
satisfy
\begin{align}
\Dslash_{\! S^n} \chi_\ell^\pm  =
\mp i \lambda_\ell \chi_\ell^\pm = \mp \textstyle\frac{i}{R} 
(\ell + \textstyle\frac{n}2) \chi_\ell^\pm \,.  \qquad (\ell \geqslant 0)
\end{align}
For $n=3$, $\lambda_\ell = \frac{i}{R}(\ell+\frac{3}{2})$ and the
spinors $\chi_{||\ell}^\pm$ transform in the
$(\frac{\ell+1}{2},\frac{\ell}{2})$ and
$(\frac{\ell}{2},\frac{\ell+1}{2})$ of $SO(4)$ which rotates the
$S^3$.  Recall that $\chi_{||}$ is that part of the spinor
$\chi=\chi_{||} \otimes \chi_{\perp}$ which is parallel to the
$S^3$.

Substituting everything back into (\ref{Diraceqn}), we obtain the
Dirac equation
\begin{align} \label{DiraceqnAdS}
 (\Dslash_{\!AdS_5} \mp \textstyle\frac{1}{R} (\ell + \textstyle\frac{3}2) - 
\textstyle\frac{1}{R}) 
\Psi^\pm_\ell  =\left\{
 \begin{matrix} 
  (\Dslash_{\!AdS_5} - \frac{1}{R} (\ell + \frac{5}2)) \Psi^+_\ell 
     \smallskip \\
  (\Dslash_{\!AdS_5} + \frac{1}{R} (\ell + \frac{1}2)) \Psi^-_\ell 
 \end{matrix}  \,
\right\} =0\,.
\end{align}
We note that the effect of the RR five-form field is to shift the
Kaluza-Klein mass by one unit. The same shift has been observed in the
dilatino spectrum on $AdS_5 \times S^5$ \cite{Kim}.
Eq.~(\ref{DiraceqnAdS}) is basically a Dirac equation on $AdS_5$ which
describes the fluctuation modes $\Psi^+_\ell$ and $\Psi^-_\ell$ with
masses
\begin{align}
 m^+_\ell = \textstyle\frac{5}{2} + \ell \,,
\qquad m^-_\ell = -(\textstyle\frac{1}{2} + \ell) \,, \label{massmode}
\end{align}
respectively. These masses are exactly the ones expected from the
field theory for fermionic bulk fields dual to the operators ${\cal
G}^\ell_\alpha$ and ${\cal F}^\ell_\alpha$,
cf. Eq.~(\ref{masses}). Moreover, the $SO(4)$ and $U(1)_R$ quantum
numbers of $\Psi_\ell^+$ and $\Psi_\ell^-$,
$(\frac{\ell+1}{2},\frac{\ell}{2})_1$ and
$(\frac{\ell}{2},\frac{\ell+1}{2})_1$, agree with those of ${\cal
G}^\ell_\alpha$ and ${\cal F}^\ell_\alpha$. This completes the
dictionary between the fermionic operators ${\cal G}^\ell_\alpha$ and
${\cal F}^\ell_\alpha$ and their dual fluctuation modes
$\Psi_\ell^\pm$.

\subsection{Dirac equation in $AdS_{d+1}$ spaces}
\label{sec32}

For the computation of the spectrum of these operators, it turns out
to be convenient to transform the Dirac equation (\ref{DiraceqnAdS})
into a second order differential equation. For this, we briefly review
Dirac equations on $d+1$-dimensional AdS spaces. 

Consider a $d+1$-dimensional AdS geometry with metric given by
\begin{align} \label{metric}
ds^2 = \frac{R^2}{z^2}  (\eta_{\mu\nu} dx^\mu dx^\nu- dz^2 )  \,,
\end{align}
where $R$ is the AdS radius. The Dirac equation for a massive
spin-$\frac{1}{2}$ mode $\Psi(x^\mu,z)$ on $AdS_{d+1}$ is then given by
\begin{align} \label{Dirac}
(  \Dslash_{AdS} -  m ) \Psi(x^\mu,z) = 0 \,,
\end{align}
where $\Dslash_{AdS} = e^M_A \gamma^A D_M$ is the Dirac operator and
$\gamma^A$ are the Dirac matrices of $d+1$-dimensional Minkowski
space, $\{\gamma^A, \gamma^B\}=2 \eta^{AB}$. The curved-space
covariant derivative
\begin{align}
D_M=\partial_M + \frac{1}{4} \omega_{MBC} [\gamma^B, \gamma^C] \,
\end{align}
is determined by the spin connection $\omega_{MBC}$ which in turn is
given in terms of the \mbox{$d+1$}-bein $e^A_M$ \cite{Sundrum:1998sj}. For an
$AdS_{d+1}$ space, $e^A_M=\frac{R}{z}$ and the Dirac operator
simplifies to \mbox{$\Dslash= \frac z{R} \gamma^A \partial_A - \frac{d}{2R}
\gamma^z$}~\cite{Henningson}. The matrix $\gamma^z$ is the
higher-dimensional analog of the chirality operator $\gamma^5$ in
$d=4$ dimensions. Multiplying the Dirac equation with $(\Dslash_{AdS}+m)$,
one obtains the following second order differential
equation~\cite{Muck}:
\begin{align} \label{KG2}
( z^2 \partial^M \partial_M - d z \partial_z - m^2 R^2 
 + \frac{d^2}{4}+\frac d 2 + m R \gamma^z ) \Psi(x^\mu,z) = 0 \,. 
\end{align}

Similarly to the case of mesons, we are interested in
spin-$\frac{1}{2}$ fluctuations which are plane waves along the
four-momentum $P_\mu$. As before, $M^2=P^\mu P_\mu$ is interpreted as
the mass of the dual spin-$\frac{1}{2}$ operator. Using the ansatz
$\tilde \Psi(x,z)= e^{iP^\mu x_\mu} f(z)$, we find
\begin{align} \label{KG} 
( z^2 \partial_z^2 - d z \partial_z + z^2 M^2 
- m^2R^2 + \frac{d^2}{4} +\frac d 2
+ mR \gamma^z ) f(z) = 0 \,. 
\end{align}
The solution of this equation~\cite{Muck} is given by
\begin{align} \label{soln}
 f(z) = z^{\frac{d+1}2} \left( J_{m-\frac1 2} (z M)\, a^+  
      + J_{m+\frac1 2} (z M)\, a^- \right)
\end{align}
where the spinors satisfy $\gamma^z a^\pm = \pm a^\pm$ and $a^- =
\frac{\gamma^\mu P_\mu}{P}\, a^+$. For small $z$, the Bessel
function $J_{m-\frac1 2}$ behaves as $J_{m-\frac1 2} (z) \approx
z^{m-\frac1 2}$, and $f(z)$ scales like $z^\Delta$.

Eq.~(\ref{KG}) is all what we will need to compute the fluctuation
spectrum in the next section. It may, however, be interesting to
extend the analysis to backgrounds which are only {\em asymptotically}
AdS. The Dirac equation on asymptotic AdS spaces is studied in
App.~\ref{secA}. Such an equation could become important for
nonsupersymmetric supergravity backgrounds.

\subsection{Spectrum of spin-$\frac{1}{2}$ fluctuations in $AdS_5$}

We now move on to compute the spectrum of the spin-$\frac{1}{2}$ modes
$\Psi^\pm_\ell$. As shown in Sec.~\ref{sec31}, these modes are
described by a Dirac equation on $AdS_5$, Eq.~(\ref{DiraceqnAdS}),
which follows from the fermionic part of the D7-brane action. In
Sec.~\ref{sec32}, this equation has been transformed into a second
order differential equation, Eq.~(\ref{KG2}). The assumption of
plane-wave fluctuations, $\Psi_\ell(x, r)=e^{iP_\mu x^\mu} f_\ell(r)$,
led then to the differential equation (\ref{KG}) which we recast into
the form ($d=4$)
\begin{align} \label{KG4}
\big( \partial_r^2 +  \frac{6}{r} \partial_r + \frac{1}{r^2} 
( -|m_\ell|^2 R^2 
 + 6 + |m_\ell| R \gamma^r )+ \frac{M^2 R^4}{r^4} \big) 
f_\ell(r) = 0 \,, 
\end{align}
where $r=R^2/z$ and the masses $m_\ell=m^\pm_\ell$ as in
Eq.~(\ref{massmode}). Without loss of generality, we work with the
absolute value of $m_\ell$, since $\gamma^r=\pm 1$. 

The above equation of motion has been derived for overlapping D3 and
D7-branes corresponding to massless quarks. We have not found a
completely satisfying way to introduce a quark mass $m_q \neq 0$ right
from the beginning. Nevertheless, it is possible to compute the mass
spectrum in dependence of the quark mass by making the following
assumption: Note that the equation for the fermionic operators,
Eq.~(\ref{KG4}), has a similar structure as the equation for the
scalar mesons, Eq.~(\ref{mesoneqnKru}). We may therefore assume that
massive quarks can be introduced by replacing
\begin{align} \label{replacement}
 \frac{M^2}{r^4} \rightarrow \frac{M^2}{(r^2+L^2)^2} \,
\end{align}
in (\ref{KG4}), where $L$ is proportional to the quark mass $m_q$,
$m_q \sim L$. This assumption will be justified below.

The resulting equation of motion 
\begin{align} 
\left( \partial_r^2 +  \frac{6}{r} \partial_r + \frac{1}{r^2} 
( -|m_\ell|^2 R^2 
 + 6 + |m_\ell| R \gamma^r )+ \frac{M^2 R^4}{(r^2+L^2)^2} \right) 
f_\ell(r) = 0 \,,
\end{align}
can then be solved in terms of the hypergeometric function ${}_2F{}_1
(a, b; c; -\frac{r^2}{L^2})$, where $a,b,$ and $c$ are constants. More
precisely, we find the solutions
\begin{align}
f_\ell(r) &= r^{|m_\ell|-3} (L^2+r^2)^{\frac{1}{2}-|m_\ell|-n_+}\,
{}_2F{}_1 \left(\textstyle\frac{1}{2}-|m_\ell|-n_+, -n_+; 
|m_\ell|+\textstyle\frac{1}{2}; -\frac{r^2}{L^2}\right) a^+ \nonumber \\
&\,\,+r^{|m_\ell|-2} (L^2+r^2)^{-\frac{1}{2}-|m_\ell|-n_-}\,
{}_2F{}_1 \left(-\textstyle\frac{1}{2}-|m_\ell|-n_-, -n_-; 
|m_\ell|+\textstyle\frac{3}{2}; -\frac{r^2}{L^2}\right) a^- \,,
\end{align}
where, as in Eq.~(\ref{soln}), the spinors $a^\pm$ satisfy $\gamma^r
a^\pm = \pm a^\pm$.  Here we defined the excitation numbers $n_\pm
=0,1,2,...$ as functions of the mass $M$ by imposing the quantization
conditions
\begin{align}
-n_+&=|m_\ell|-\textstyle\frac{1}{2}\sqrt{1+M^2/L^2} \,,  \label{n+}
 \\
-n_-&=|m_\ell|+1-\textstyle\frac{1}{2}\sqrt{1+M^2/L^2} \,.\label{n-}
\end{align} 
These conditions ensure that the hypergeometric functions behave like
$(r^2)^{n_\pm}$ at \mbox{$r \rightarrow \infty$}, and thus asymptotically
$f_\ell(r) \sim r^{-\Delta}$. 
Since the fluctuations $\Psi_\ell^\pm$ are canonically normalized,
this is the expected asymptotic behavior for normalizable modes. 

The quantization conditions (\ref{n+}) and (\ref{n-}) determine the
mass spectra of the operators ${\cal G}^\ell_\alpha$ and ${\cal
F}^\ell_\alpha$. The function $f_\ell$ is a superposition of two
solutions which are proportional to $a^+$ and $a^-$.  For
\mbox{$m_\ell=m_\ell^+>0$}, the dominant solution at $r \rightarrow
\infty$ is that proportional to $a^+$. Solving the definition of $n_+$
for $M$ and substituting $m_\ell^+= \frac{5}{2}+\ell$, we obtain
\begin{align} \label{massspectrum}
M^2_{\cal G}&=\frac{4L^2}{R^4} (|m_\ell^+|+n_+ -\textstyle\frac{1}{2})
(|m_\ell^+|+n_+ +\textstyle\frac{1}{2})  \nonumber\\
&= \frac{4L^2}{R^4} (n_+ +\ell+2)
 (n_+ +\ell+3) \qquad (n_+ \geqslant 0, \ell \geqslant 0)
\,.
\end{align}
This is the mass spectrum of the operators ${\cal G}^\ell_\alpha$
which have the same $SO(4)$ quantum numbers as $\chi^+$,
$(\frac{\ell+1}{2}, \frac{\ell}2)$.

For \mbox{$m_\ell=m_\ell^-<0$}, the two solutions interchange their
roles. Inverting now the definition of $n_-$ and substituting
$|m_\ell^-| = \frac{1}{2} + \ell$, we obtain the mass spectrum of
${\cal F}_\alpha^\ell$:
\begin{align}
 M^2_{\cal F}&=\frac{4L^2}{R^4} (|m_\ell^- | +n_- +\textstyle\frac{1}{2})
(|m_\ell^-|+n_- + \textstyle\frac{3}{2})  \nonumber\\
&= \frac{4L^2}{R^4} (n_- + \ell+1)
 (n_- + \ell+2) \qquad (n_- \geqslant 0, \ell \geqslant 0) \,.
\end{align}
The mass spectrum transforms as $\chi^-$ in the $(\frac{\ell}{2},
\frac{\ell+1}2)$ of $SO(4)$.

\subsection{Fluctuation-operator matching}

In the following we summarize all fluctuations of the D7-brane and
assign the corresponding operators to them.  As was found in
\cite{Kruczenski}, the complete set of open string fluctuations fits
into a series of massive supermultiplets of the $\N=2$ supersymmetry
algebra. These multiplets have the masses
\begin{align}
  M^2 = \frac{4L^2}{R^4} (n+ \ell +1)(n+\ell+2) \qquad (n,\ell
  \geqslant 0) \,
\end{align}
and are labeled by the quantum number $\ell$. Since the supercharges
commute with the generators of the global group $SU(2)_\Phi$,
the $SU(2)_\Phi$ quantum number, $\frac{\ell}{2}$ is the same for
all fluctuations in a supermultiplet. 

The bosonic fluctuations of a multiplet are listed in the upper part
of Tab.~\ref{tableop}. The notation of the fluctuations and their mass
spectra is the same as in \cite{Kruczenski}. The numbers $(j_1,j_2)_q$
label a representation of $SO(4) \approx SU(2)_\Phi \times SU(2)_{\cal
R}$, and $q$ is the $U(1)_R$ charge. The last column shows the
conformal dimension of the lowest operator in a series.

Let us consider the bosonic components in more detail. First, there is
a scalar in the ($\frac{\ell}{2}$, $\frac{\ell}{2}+1$)$_0$ which 
corresponds to the chiral primaries \cite{Kruczenski}
\begin{align}\label{primaries}
 {\cal C}^{I\ell}= \bar q^m \sigma_{mn}^I X^\ell q^n \,,
\end{align}
where the Pauli matrices $\sigma_{mn}^I$ ($I=1,2,3$) transform in
triplet representation of $SU(2)_R$. Then, there are 2 scalars in the
($\frac{\ell}{2}$, $\frac{\ell}{2}$)$_2$ which we identified in
Sec.~\ref{sec22} as the scalar meson operators ${\cal M}_s^{A\ell}$
($A=1,2$). Moreover, there is 1 scalar and 1 vector\footnote{These
fields fit into a 5d vector field of $AdS_5$.} in the
($\frac{\ell}{2}$, $\frac{\ell}{2}$)$_0$ which we identify as the
operators
\begin{align}
 {\cal J}^{\mu\ell}_B = \bar \psi_i^\alpha \gamma^\mu_{\alpha\beta}
X^\ell \psi_i^\beta + i \bar q^m X^\ell D^\mu q^m
+ i \bar  D^\mu \bar q^m X^\ell q^m \qquad (\mu=0,1,2,3)
 \,,
\end{align}
with $X^\ell$ as in Eq.~(\ref{mesonop}). The operator ${\cal
J}^{\mu0}_B$ at the bottom of the tower is associated with the global
$U(1)_B$ current. Finally, there is a scalar in the ($\frac{\ell}{2}$,
$\frac{\ell}{2}+1$)$_0$ ($\ell\geq 2$) which is a higher descendant of
${\cal C}^{I\ell}$. 

There are analogous operators in the defect conformal field theories
located on the T-dual D3/D5 and D3/D3 brane intersections
\cite{0111135, Constable}. For instance, the operator ${\cal
M}_s^{A\ell}$ ($A=1,2$) corresponds to the scalar mesons
${\cal E}^{A\ell}$ ($A=1,2,3$) \cite{0111135} and ${\cal C}^{\mu\ell}$
($\mu=1,2,3,4$) \cite{Constable} in the D3/D5 and D3/D3 theories,
respectively.  Also, the $U(1)_B$ current operator ${\cal J}^{\mu0}_B$
is part of the lowest multiplet in each of the D3/D$p$ ($p=3,5,7$)
theories.   

The fermionic content of the multiplets is shown in the lower part of
Tab.~\ref{tableop}. This part of the multiplet matches precisely the
spectra of the fermionic operators:\footnote{The mass spectra
$M^2_{\cal G}$ and $M^2_{\cal F}$ are identical to the spectra
$M^2_{F2}$ and $M^2_{F1}$ found in \cite{Kruczenski}, respectively.}
\begin{align}
M_{\cal G}(n, \ell-1)=M_{\cal F}(n, \ell) \qquad
(\ell \geqslant 1) \,.
\end{align}
The lowest multiplet with $\ell=0$ contains only the operator ${\cal
F}_\alpha^0$. The operator ${\cal G}_\alpha^\ell$ appears first for
$\ell=1$. We have already discussed the structure of the operators
${\cal F}_\alpha^\ell$ and ${\cal G}_\alpha^\ell$ in Sec.~\ref{sec31}.

\begin{table}[ht] 
\begin{center}
\begin{tabular}{cccccccc}
      &\!\! fluctuation & d.o.f. & $(j_1,j_2)_q$ & spectrum && op. & $\Delta$\\
\hline
bosons & 1 scalar  &1& ($\frac{\ell}{2}$, $\frac{\ell}{2}+ 1$)$_0$ &  $M_{I,-}(n,\ell+1)$  & $(\ell \geq 0)$ & ${\cal C}^{I\ell}$ & $2$ \\
       & 2 scalars &2& ($\frac{\ell}{2}$, $\frac{\ell}{2}$)$_2$    & $M_s(n,\ell)$ & $(\ell \geq 0)$ & ${\cal M}_s^{A\ell}$  & $3$\\
       & 1 scalar  &1& ($\frac{\ell}{2}$, $\frac{\ell}{2}$)$_0$    & $M_{III}(n,\ell)$  & $(\ell \geq 1)$ & \raisebox{-0.2cm}{${\cal J}^{\mu\ell}_B$} \vspace{-0.14cm} & \raisebox{-0.2cm}{$3$}\\
       & 1 vector  &3 & ($\frac{\ell}{2}$, $\frac{\ell}{2}$)$_0$    & $M_{II}(n,\ell)$  & $(\ell \geq 0)$ & \\
       & 1 scalar  &1& ($\frac{\ell}{2}$, $\frac{\ell}{2}- 1$)$_0$ &  $M_{I,+}(n,\ell-1)$  & $(\ell \geq 2)$ & -- & $4$  \\
\hline
fermions  & 1 Dirac  &4 & ($\frac{\ell}{2}$, $\frac{\ell + 1}{2}$)$_1$ & $M_{{\cal F}}(n, \ell)$  & $(\ell \geq 0)$& ${\cal F}^\ell_\alpha$ & $\frac{5}{2}$ \\
  & 1 Dirac &4& ($\frac{\ell}{2}$, $\frac{\ell - 1}{2}$)$_1$ & $M_{{\cal G}}(n, \ell-1)$  & $(\ell \geq 1)$ & ${\cal G}^\ell_\alpha$ & $\frac{9}{2}$
\end{tabular}
\end{center} 
\caption{Field content of supermultiplets in the D3/D7 theory.}\label{tableop}
\end{table}

As required by supersymmetry, the number of bosonic components in a
multiplet,
\begin{align}
1 (2 (\textstyle\frac{\ell}2+1)+1) + (2+1+3)(2 \textstyle\frac{\ell}2+1) +  
1 (2 (\textstyle\frac{\ell}2-1)+1) = 8(\ell+1) \,,
\end{align}
agrees with the number of fermionic components,
\begin{align}
4 (2 \textstyle\frac{\ell+1}2+1) +  
4 (2 \textstyle\frac{\ell-1}2+1) = 8(\ell+1) \,.
\end{align}

The masses of the fermionic fluctuations matches exactly the spectrum
expected from supersymmetry. With hindsight, this justifies the
introduction of the quark mass into the equations of motion via the
replacement (\ref{replacement}).

\section{Baryon spectroscopy in the ``bottom-up'' approach}
\label{sec4}

So far we have discussed the spectrum of bosonic and fermionic
operators with two fundamental fields at the ends. In contrast, a
baryon in large $N$ $SU(N)$ (super) Yang-Mills theory is a color
singlet bound state of $N$ fundamental quarks. As discussed in the
introduction, the construction of a brane configuration for dynamical
baryons is remarkably difficult.  Nevertheless, it is possible to
derive the baryon spectrum in the so-called ``bottom-up'' approach to
the AdS/CFT correspondence.  Starting from a phenomenological
supergravity model, we will study the baryon spectrum of a broad class
of supersymmetric field theories.

\subsection{Mass spectra in the D3/D7 theory in the effective approach}
\label{sec41}

Let us first demonstrate this technique by computing once again the
scalar meson spectrum in the $\N=2$ theory of the D3/D7 configuration.
Assume we had no knowledge about the dual gravity theory apart from
its existence. We may then construct the supergravity background from
the properties of the field theory. According to the standard
prescription of AdS/CFT, the conformal invariance of the field theory
requires the dual supergravity background to be $AdS_5$.  The
$SO(4,2)$ isometry of $AdS_5$ corresponds to the conformal group of
the field theory.

Next, in order to compute the spectrum of the operators ${\cal
M}^\ell_s$ as defined in Eq.~(\ref{mesonop}), we introduce scalar modes
$\phi_\ell$ in this $AdS_5$ space which are dual to the operators
${\cal M}^\ell_s$. These scalars are described by the equation of
motion
\begin{align}
\partial_M \sqrt{g} g^{MN} \partial_N \phi_\ell - m^2_\ell \phi_\ell = 0 \,,
\end{align}
where $g_{MN}$ is the $AdS_5$ metric in the parametrization
(\ref{metric}), and $m_\ell$ is fixed by the mass-dimension relation
\begin{align} \label{mdr}
m^2_\ell = \Delta(\Delta-4) = -3 + \ell(\ell+2) \,.
\end{align}
Using again the plane-wave ansatz $\phi_\ell= e^{-ik\cdot x} f_\ell$
with $M^2=-k^2$, this leads to
\begin{align}
  (z^2 \partial_z^2 -3z \partial_z + z^2 M^2 -m_\ell^2 R^2)
  f_\ell(z) =0 \,,
\end{align}
or, equivalently ($r=R^2/z$), 
\begin{align} \label{mesoneqn}
\partial_r^2 f_\ell(r)+ \frac{3}{r} \partial_r f_\ell(r)+
\left(\frac{M^2}{r^4} - \frac{\ell(\ell+2)}{r^2}\right) f_\ell(r)
=0\,,
\end{align}
where we redefined $f_\ell \rightarrow f_\ell/r$.
After the replacement (\ref{replacement}), this differential equation
becomes identical to Eq.~(\ref{mesoneqnKru}) which we obtained from
the DBI action of the D7-branes. The result for the meson spectrum
is therefore the same as in the full ten-dimensional string approach.

Several remarks are in order here: First, the DBI computation in Sec.\
\ref{sec2} has shown that the eigenvalues of the spherical harmonics
on $S^5$ lead to the dependence of the scalar mass $m_\ell$ on the
angular momentum $\ell$. In the phenomenological approach, the
internal space $S^5$ may be ignored in the effective approach. The
dependence on $\ell$ enters the mass via the conformal dimension of
the higher-$\ell$ operators, see Eq.~(\ref{mdr}). Second, it is not
obvious how a nonvanishing quark mass can be introduced in this
approach. The introduction of the appropriate dual field in the
background of the induced metric on the D7-brane does not lead to the
desired result. Inspection of the DBI action shows that the metric
elements transverse to the D7-brane are relevant for nonvanishing
quark mass. These metric elements do not appear in an effective
computation.  We therefore use the replacement (\ref{replacement}) for
the introduction of a quark mass. Third, we may also compute the
spectrum of the fermionic operators ${\cal F}^\ell_\alpha$ and ${\cal
G}^\ell_\alpha$ in the ``bottom-up'' approach. In this case we would
introduce a Dirac spinor $\Psi$ in $AdS_5$. Fixing the masses as in
Eq.~(\ref{masses}), we immediately obtain Eq.~(\ref{DiraceqnAdS}) from
which we obtained the spectrum for ${\cal F}^\ell_\alpha$ and ${\cal
G}^\ell_\alpha$.

\subsection{Baryons in superconformal field theories}
\label{sec42}

We now turn to baryon operators in a broad class of large $N$ $SU(N)$
super Yang-Mills theories with $N_f$ flavors. The only further
assumption we make is that the theory is conformal invariant, at least
in some parameter regime, and that baryons do exist in the theory. An
example of such a theory would be super QCD in the conformal window
($3/2 \leq N_f/N_c \leq 3$) or asymptotic free theories at high
energies. Conformal invariance ensures that the dual supergravity
background has the structure of an $AdS_5$ space.

In this class of theories, we are interested in the spectrum of the 
totally antisymmetric baryon operator
\begin{align} 
{\cal B}^0 = \frac{1}{
\sqrt{N!}} \, \varepsilon_{i_1 i_2 ... i_N} \psi_{i_1}...
 \psi_{i_N} \,
\end{align}
which has conformal dimension $\Delta=\frac{3}{2}N$ and spins
$\frac{1}{2}$,...,$\frac N 2$ ($N$ odd). For simplicity, we only
consider the state with spin $\frac 1 2$. We can also construct
operators ${\cal B}^\ell$ of higher orbital excitation by the
insertion of $\ell$ derivatives $D_{i_k}$ into ${\cal B}^0$. Such
operators would have the conformal dimensions
\begin{align}\label{deltaop}
\Delta=\frac{3}{2}N+\ell \,.
\end{align}  

Let us now specify the characteristics of supergravity fields
$\Psi_\ell$ dual to the operators~${\cal B}^\ell$. Again, the spin of
$\Psi_\ell$ must be identical to that of the boundary states which we
have chosen to be $\frac{1}{2}$.  Moreover, the conformal dimensions
of the operators ${\cal B}^\ell$ as given by Eq.~(\ref{deltaop})
determine the masses of the fields $\Psi_\ell$.  Using
Eq.~(\ref{eqn1}), we find the masses
\begin{align} 
m_\ell =  \textstyle\frac{3}{2} N - 2 + \ell 
\,. \label{massbulk}
\end{align}

In the phenomenological approach the spin-$\frac{1}{2}$ component of
the baryons ${\cal B}^\ell$ are quite similar to the fermionic
operators ${\cal G}^\ell_\alpha$ (${\cal F}^\ell_\alpha$).  The dual
fields $\Psi_\ell$ are described by a Dirac equation on $AdS_5$,
\begin{align}
(  \Dslash -  m_\ell ) \Psi_\ell(x,z) = 0 \,,
\end{align}
but now the masses $m_\ell$ are given by Eq.~(\ref{massbulk}).
Proceeding as in Sec.~\ref{sec3}, we obtain the baryon masses 
\begin{align}
  M^2_{\cal B}&= \frac{4L^2}{R^4} (m_\ell+n -\textstyle\frac{1}{2})
(m_\ell + n +\textstyle\frac{1}{2})  \qquad (n,\ell \geqslant 0) \,,
\end{align}
cf.\ Eq.~(\ref{massspectrum}). Substituting the masses $m_\ell$ as
given by Eq.~(\ref{massbulk}), we find the baryon spectrum
\begin{align}
  M^2_{\cal B} &=\frac{4L^2}{R^4} (n+\ell+ \textstyle\frac{3}{2}(N-1))
  (n+\ell+ \textstyle\frac{3}{2}(N-1)-1) \,. \label{baryonspectrum}
\end{align}
For large $N$, and constant $n$ and $\ell$, the baryon
masses $M_{\cal B}$ scale with $N$, as expected from field theory
\cite{Witten:1979kh}.

\section{Conclusions and open problems}

We have derived the mass spectra of certain fermionic operators in the
$\N=2$ field theory located on the world-volume of intersecting D3 and
D7-branes. This supplements the \mbox{analysis} of \cite{Kruczenski}
in which the spectrum of scalar and vector meson operators was found
in the same theory. We showed that both the bosonic as well as the
fermionic operators fit into $\N=2$ supermultiplets and constructed
explicit expressions for these operators. We finally made a prediction
for the mass of baryon operators in a class of supersymmetric field
theories using an effective approach to the AdS/CFT correspondence.

In the derivation of the fermionic operator spectrum we made the
assumption that massive quarks can be introduced via the replacement
(\ref{replacement}). The procedure led  exactly to the operator
spectrum expected from supersymmetry, which justifies the replacement
with hindsight. Certainly, it would be more desirable to include a
nonvanishing quark mass already on the level of the brane set-up.
Another goal would be to extend the analysis to nonsupersymmetric and
nonconformal backgrounds. In this case the masses of the bosonic
fluctuations would differ from that of the fermionic ones. 

Finally, one purpose of this paper was to develop techniques which
will also be relevant for the holographic discussion of dynamical
baryons. We mainly focused on open string fluctuations dual to
operators with half-integer spin. As outlined in the introduction, it
would be interesting to find a 10d baryon configuration and compute
the corresponding operator spectrum. If the brane configuration turns
out to be of the form of an $AdS$ space, at least in some parameter
regime, the baryon spectrum should agree with that found in the
effective approach. It would be nice, if the spectrum
(\ref{baryonspectrum}) could be verified from the full 10d string
theory point of view.

%
\section*{Acknowledgments}
%
I would like to thank N.~Arkani-Hamed, C.~Beasley, J.~Erdmenger,
Z.~Guralnik, and L.~Motl for helpful discussions related to this work.
I also thank C.~Nu$\tilde {\rm n}$ez for numerous comments on the
first version of the paper. The author was supported by a fellowship
within the Postdoc-Program of the German Research Society (DFG), grant
KI~1084/1.

\appendix

\section{Dirac equation in asymptotic $AdS_{d+1}$ spaces}
\label{secA}

In this section we study the Dirac equation on asymptotic $AdS_{d+1}$
spaces. The general class of backgrounds we wish to consider is
given by the metric and dilaton
\begin{align} \label{metricasymp}
ds^2 = \frac{R^2}{z^2} e^{2A(z)} (\eta_{\mu\nu} dx^\mu dx^\nu
- dz^2 )  \,, \qquad \Phi=\Phi(z)\,,
\end{align}
where $A(z), \Phi(z) \rightarrow 0$ as $z=R^2/r \rightarrow
0$, and $R$ is the AdS radius. 

The Dirac action for a spinor $\Psi(x^\mu,z)$ in this background is
given by
\begin{align}
S= \int d^{d} x dz\, e^{-\Phi} \sqrt{-\det g} \left[ \bar \Psi e^M_A\gamma^A (
D_M- \frac{1}{2} \partial_M \Phi) \Psi - m \bar \Psi \Psi \right] \,,
\end{align}
where $x^M=(x^\mu,z)$. The resulting Dirac equation is
\begin{align}
(  \Dslash - \frac{1}{2} e^M_A \gamma^A\partial_M \Phi -  m ) \Psi(x^\mu,z) = 0 \,,
\end{align}
again with $\Dslash = e^M_A \gamma^A D_M$. The dependence on the
dilaton can be separated by the ansatz
\begin{align}
\Psi(x^\mu,z) = e^{\frac 1 2 \Phi(z)} \tilde \Psi(x^\mu,z) \,,
\end{align}
and we just have to solve $(\Dslash -m)\tilde \Psi=0$.  If $\Phi(z)$
is a regular function, i.e.~$e^{\frac{1}{2}\Phi(z)}>0$ for all $z$,
then $\Psi$ has the same zeros as $\tilde \Psi$.

For the background (\ref{metricasymp}), the $d+1$-bein is $e^A_M=\delta^A_M
a(z)$ with $a(z) \equiv \frac{R}{z} e^{A(z)}$.  The only non-vanishing
component of $\omega_{MAB}$ is $\omega_{\mu a z}=\eta_{\mu a}
\frac{\partial_z a(z)}{a(z)}$ \cite{Contino:2004vy} and the Dirac
operator becomes
\begin{align}
\Dslash = a(z)^{-1} \left[\gamma^A \partial_A 
+ \frac{d}{2}  b(z) \gamma^z\right] \,,\qquad b(z) \equiv
\partial_z \ln a(z)  \,.
\end{align}
It is then straight-forward to get
\begin{align} \label{KGasymp}
( \partial_z^2 + d\, b(z)\, \partial_z + M^2 
- a(z)^2 m^2 + \frac{d^2}{4}\, b(z)^2 
+ \frac d 2\, \partial_z b(z)
- \partial_z a(z) \gamma^z m ) f(z) = 0 \,.
\end{align}
This equation reduces to Eq.~(\ref{KG}) for $a(z)=\frac{R}{z}$
($A(z)=0$).  It can be used for the computation of the spectrum of
spin-$\frac{1}{2}$ fluctuations in confining backgrounds with
non-trivial warp factor. We do not pursue this issue any further here.

\end{document}